\documentclass[aps,pra,10pt,twocolumn,showpacs,superscriptaddress,nofootinbib,preprintnumbers]{revtex4-2}

\usepackage[dvipsnames]{xcolor}
\usepackage{hyperref}
\hypersetup{
  breaklinks=true,
  colorlinks=true,
  allcolors=BlueViolet,
} 
\usepackage{url}

\usepackage{physics}
\usepackage{amsmath}
\usepackage{enumitem}
\usepackage{comment}

\usepackage{amsthm}
\usepackage{bm}% bold math
\usepackage{graphicx}% Include figure files
\graphicspath{{./figures/}} % set path for all figures

\usepackage{braket}
\usepackage{amsmath}

\newcommand{\D}{\mathcal{D}}

\renewcommand{\O}{\mathcal{O}}

\usepackage{dsfont}

\begin{document}

\preprint{}

\title{Quantum Fisher Information and the Curvature of Entanglement}

\author{Zain H. Saleem}
\affiliation{Mathematics and Computer Science Division, Argonne National Laboratory, Lemont, IL, USA}

\author{Anil Shaji}
\affiliation{School of Physics, IISER Thiruvananthapuram, Kerala, India 695551}

\author{Anjala M Babu}
\affiliation{School of Physics and Applied Physics, Southern Illinois University, Carbondale, IL, USA}

\author{Da-Wei Luo}
\affiliation{Center for Quantum Science and Engineering and Department of Physics, Stevens Institute of Technology, Hoboken, New Jersey 07030, USA}

\author{Quinn Langfitt}
\affiliation{Department of Computer Science, Northwestern University,
Evanston, IL, U.S.A.}

\author{Ting Yu}
\affiliation{Center for Quantum Science and Engineering and Department of Physics, Stevens Institute of Technology, Hoboken, New Jersey 07030, USA}

\author{Stephen K. Gray}
\affiliation{Center for Nanoscale Materials, Argonne National Laboratory, Lemont, IL, USA}

\begin{abstract}
We explore the relationship between quantum Fisher information (QFI) and the negative of the second derivative of concurrence with respect to the coupling between two qubits, referred to as the curvature of entanglement (CoE). The two-qubit system serves as a minimal model to study the connection between QFI and dynamically generated entanglement in scenarios where the measured quantity is a two- or many-body coupling strength.  We analyze in detail the pure-state lossless case for which general results can be inferred and we also consider a simple interaction Hamiltonian in the case of one form of loss applied to the qubits. For a two-qubit quantum probe used to estimate the coupling constant appearing in the interaction Hamiltonian we show, for certain initial conditions, that there are times such that CoE = QFI. These times can be associated with the concurrence, viewed as a function of the coupling parameter, being a maximum. We examine the time evolution of the concurrence of the eigenstates of the symmetric logarithmic derivative (SLD). Measurements using the SLD eigenstates as basis are optimal for saturating the quantum Cramer bound.  We show that, for several families of initially separable and initially entangled states, the SLD eigenstates are simple product states when CoE = QFI. 

%simple product measurements suffice to \textcolor{purple}{when CoE = QFI, while otherwise, in general, entangled measurements are required
%\textcolor{teal}{if one wishes or it is desirable to use SLD eigenstates for the projective measurements.}
%giving an operational significance to the points in time when CoE = QFI. 
%\textcolor{purple}{We also prove that for generic two-qubit pure states CoE $\leq$ QFI.} \textcolor{blue}{We dont include the single qubit terms so the state is not completely generic. we should clarify that this is only for interacting two-qubit hamiltonians?} (\textcolor{purple}{The states are generic since all we do is to take a random pure state and expand it in the Bell basis in Eqs (9) and (10). The Hamiltonian is not generic since we assume that its eigenstates are Bell states which is true only if the Hamiltonian has the specific Bell-diagonal form that we use.})
%\textcolor{teal}{For the pure-state case we also prove, for general two-qubit nonlocal Hamiltonians  and all possible initial conditions, that CoE $\leq$ QFI.}

%\textcolor{teal}{\textbf{SKG}: given that we know, for some cases, simple $|10>, |01>$ measurements work for all g, I think we need to be cautious and so I toned down abstract  relative to previous.  It's still a really deep result as it stands. Also, since there is not much published clarity out there on SLD eigenstates being better, from a practical standpoint,  I toned down a bit within the text.  See also some comments starting with SKG both within the pdf and the tex file.}

\end{abstract}
\maketitle

\section{Introduction}

The quantum Fisher information (QFI)~\cite{Helstrom1976,Holevo2011,Braunstein1994} has emerged as a central quantity at the intersection of quantum metrology, quantum information, and condensed matter physics. The QFI, in the context of quantum metrology, bounds the precision with which a parameter can be estimated through measurements using a quantum probe~\cite{Paris2009}. In its condensed physics setting, QFI has been shown to signal phase transitions~\cite{ZanardiParis2008}. The QFI can also be connected directly to experimentally accessible quantities like dynamical structure factors measured using by neutron scattering, providing an operational witness of 
multipartite entanglement in correlated quantum matter~\cite{Hauke2016}.

The framework of quantum-enhanced metrology casts entanglement as a static resource associated with the initial state of the quantum probe. It becomes a resource that is to be engineered into the probe state before the parameter-encoding evolution begins~\cite{Giovannetti2004,Giovannetti2006}. If the quantum probe is assumed to be made up of $N$ qubits, then the parameter dependent evolution of the probe is generated by a local Hamiltonian of the form  
\begin{equation}
    \label{Hsep1}
    H_{\rm probe} = \theta\sum_i h_i,
\end{equation} 
where $h_i$ are operators on individual qubits. For this non-entangling evolution, the right kind of entanglement engineered into the initial state is the enabling factor that allows the measurement protocol to saturate the quantum Cram{\'e}r--Rao bound~\cite{Braunstein1994} and achieve a $1/N$ scaling for the measurement precision as opposed to the classical, shot-noise limited, scaling of $1/\sqrt{N}$. The role of initial multipartite entanglement and non-classical correlations as a resource quantum metrology is now well established~\cite{Hyllus2012, Toth2012, Toth2014, PezzeRMP2018}.

In the context of many-body systems, quantum phase transitions were identified as points where there is maximum sensitivity to the parameter $\theta$~\cite{Zanardi2007geo, VenutiZanardi2007, ZanardiParis2008}. Quantum critical systems can therefore be considered as a resource for parameter estimation and it was shown that the divergence of the fidelity susceptibility at a critical point translates into an improvement in precision of the order of the inverse of the system size compared to the non-critical regime. Fidelity susceptibility was, in turn, connected to the dynamical structure factor~\cite{YouLiGu2007} providing a connection between metrological advantage and spectroscopically measurable quantities. This line of investigations led to the connection between the dynamic susceptibility and QFI density obtained in~\cite{Hauke2016}.

It is worth noting that even though the natural Hamiltonian for the condensed matter systems considered consists of many-body couplings the parameter estimated using critical systems and the parameter corresponding to the QFI density extracted from dynamical susceptibility are all associated with single-qubit terms as in Eq.~\eqref{Hsep1}. In these cases the QFI plays the role of a witness for genuine multipartite entanglement in the initial state of the probe which is the equilibrium state of the many-body system.

The case in which the parameter of interest is itself an entangling two-body coupling, and the probe entanglement is dynamically generated by the same interaction that encodes the parameter has received considerably less systematic attention. One study we are aware of is on determination of the coupling in the Jaynes-Cummings Hamiltonian \cite{genoni2012}, although it does not focus on dynamical evolution and entanglement as we do here.

Entanglement measures are functions only of the instantaneous state of the probe while the QFI depends both on the initial state and the dynamics making it challenging to establish such connections. Here we address this issue by investigating the relationship between QFI and the parametric derivatives of entanglement measures thus introducing a parametric dependence on both sides of the relationship. The entanglement measure that we will study in this work is the concurrence \cite{wootters1998entanglement} which is an effective measure for quantifying bi-partite entanglement. General results for the lossless
case are obtained and we also consider some cases with identical and independent losses on the two qubits. 
We find quite remarkable connections between the time-dependence of the QFI and the the negative of the second derivative of the concurrence with respect to the measured parameter (which we term the curvature of entanglement or CoE) when the measured quantity is in itself the coupling parameter. The negative sign in our definition of CoE arises because the points in time where the concurrence of the state of the two-qubit probe is maximal are of particular interest to us. The second derivative of concurrence is negative at these maximas and the additional negative sign in our definition of CoE compensates for this making CoE directly comparable to QFI which is always greater than zero. In fact, there are several other interconnected relations concerning the concurrence and measurements that we detail. We should note that the derivatives of concurrence with respect to coupling and its analogues have previously been studied in a different context in relation to phase transitions in many body systems \cite{osterloh2002scaling,amico2004dynamics,wu2004quantum,yang2005reexamination}. Most these
studies focused on the first derivative although Ref. \onlinecite{osterloh2002scaling} did
present some analysis of the second derivative. However we see no direct relation of these latter works
to our own.

Section II below gives brief definitions of the quantities of interest, including an elaboration on the nature of the two-qubit interaction Hamiltonian.  Sec.~III presents analytical results for the types of Hamiltonian and initial conditions for which the relations hold in the pure-state, noise-free limit. 
In Sec. IIIB, an inequality, CoE $\leq$ QFI, is shown to be true for general nonlocal interaction
Hamiltonians and all initial conditions in the pure-state, noise-free limit.
An illustrative example also including amplitude loss is given in Sec.~IV, Appendices give additional results, and Sec.~V presents concluding remarks.

{\section{ QFI, Two-Qubit Hamiltonian and Concurrence}

Consider the problem of estimating a scalar parameter $\theta$ using a quantum probe, with an (unbiased) estimator denoted by $\hat{\theta}$. The parameter-dependent evolution of the probe is generated by the  Hamiltonian $H_{\rm probe} = \theta h$ of the form in Eq.~\eqref{Hsep1} but now without the restriction that $h$ is a sum of single qubit operators.  The theoretical lower limit on the measurement uncertainty is given by the quantum Cramér-Rao bound (QCRB) \cite{holevo2011probabilistic},
\begin{align}
  \delta \hat \theta ^2 \geq \frac{1}{M F(\theta)},
\end{align}
where $M$ is the number of independent applications of the quantum probe and $F(\theta)$ is the QFI.

For a generic initial probe state $\rho = \sum_a \lambda_a \op{\lambda_a}$ written in its eigenbasis, the QFI with respect to the parameter $\theta$ is given by \cite{liu2019quantum}:
\begin{align}
    \label{fisher1}
F(\theta) = 2 \sum_{a,b}
\frac{\abs{\bra{\lambda_a} (\partial_\theta \rho ) \ket{\lambda_b}}^2}
{\lambda_a + \lambda_b},
\end{align}
where the summation is restricted to indices $a, b$ such that $\lambda_a + \lambda_b \neq 0$ and $\partial_\theta \rho$ represents the derivative of $\rho$ with respect to the parameter $\theta$. The derivative,  $\partial_\theta \rho$, can be obtained as the action of the \textit{symmetric logarithmic derivative} (SLD) operator on $\rho$. The SLD, denoted by $L$, is defined implicitly via $\partial_\theta \rho = ( L \rho + \rho L )/2$. In terms of $L$ the QFI can be succinctly expressed as $F(\theta) = \mathrm{Tr}[\rho L^2]$. The SLD and through it, the QFI depends on the generator $h$ of translations in $\theta$ as well as the initial state of the quantum probe.

When designing an implementation of quantum limited metrology, in addition to choosing an initial probe state that maximizes the QFI, it is important to choose a read-out strategy on the probe that provides maximum information about $\theta$. The eigenbasis of the SLD operator provides the optimal read-out basis that saturates the QCRB \cite{Braunstein1994, braunstein1996generalized, Toth2014}. However, typically, the eigenstates of $L$ are entangled states of the $N$ probe units and projective measurements onto a set of entangled states are not easy to implement in most cases. In this paper we also look at the entanglement in the SLD eigenstates corresponding to instantaneous probe states which gives us additional clues about the ease of implementing a measurement scheme that saturates the QCRB and its relationships to the entanglement in the probe state, the instantaneous QFI and the parametric derivatives of the concurrence of probe states. 

In the following, we will be considering quantum probes made of two interacting qubits, with the parameter estimation problem corresponding
to finding the coupling parameter 
which is often denoted as $g$, i.e. we choose $\theta$ = $g$. 
The general, nonlocal or interaction two-qubit Hamiltonian is
%The most general form for the generator $h$ in this case is $\sum_{\alpha, \beta} \eta_{\alpha, \beta} \sigma_\alpha \otimes \sigma_\beta$, $\alpha, \beta = 0, x,y,z$, where $\sigma_j$, for $j=x,y,z$ are the Pauli matrices and $\sigma_0 = \openone_2$ (the $2 \times 2$ identity matrix). The parameters $\eta_{\alpha \beta}$ in front of each of the terms are either 0 or a numerical constant since we are considering the single parameter estimation scenario only and all the terms in $H_{\rm probe}$ have to be proportional to the parameter that is to be estimated. 
%As mentioned earlier, we are specifically interested in the case where the coupling between the two qubits is estimated. This means that $h$ does not contain single qubit terms of the form $\sigma_0 \otimes \sigma_j$ or $\sigma_j \otimes \sigma_0$. So, for the two-qubit probe, the general parameter estimation problem we are interested in corresponds to 
\begin{equation}
\label{inter}
H_{\rm probe} ~= g \sum_{j,k = x,y,z} \eta_{jk} \sigma_j \otimes \sigma_k~~~,  
\end{equation}
where $\sigma_j$, for $j=x,y,z$, are the Pauli matrices and the $\eta_{jk}$ are real.
The coefficient matrix $\eta$ can be  diagonalized via its singular value decomposition as $\eta =UDV^T$, where $D$ is diagonal matrix and with proper choice of signs of the elements of $D$, we can make the both $U$ and $V$ orthogonal transformations ($U,V \in SO(3)$). The orthogonal transformations that diagonalize $\eta$ can instead be considered as redefinitions of the Pauli operators for each of the two qubits as (see Appendix A), 
\begin{align}
    h &= \sum_{j,K} \eta_{jk} \sigma_{j} \otimes \sigma_{k} 
    = \sum_{j,K} \left[U D V^T\right]_{jk} \sigma_{j} \otimes  \sigma_{k} \nonumber \\
    &= D_{ll} \left[U_{lj}^T\sigma_{j}\right] \otimes \left[V^T_{lk} \sigma_{k}\right] \nonumber \\
    &\equiv \eta_{k}\, \tilde\sigma_{k} \otimes  \tilde\sigma_{k}.
\label{rotate_h}
\end{align}
The operators $\tilde{\sigma}_k$ are redefinitions of the $\sigma_k$ operators which is equivalent to a particular choice of the $x,y$ and $z$ directions for each qubit. Through this orthogonal transformation  we see that we can bring Hamiltonian of the two-qubit probe to the $\eta$-diagonal, anisotropic Heisenberg form,
\begin{equation}
\label{HprobeF}
\widetilde{H}_{\rm probe} =g \sum_{k = x,y,z} \eta_k \, \sigma_k \otimes \sigma_k~~~. 
\end{equation}
Note that in the absence of the single qubit operators in $H_{\rm probe}$, there are no pre-defined $x,y$ and $z$ directions in space for each qubit and we are at liberty to always simplify $H_{\rm probe}$ into the form given above.  
An equivalent way of stating this result is $H_{\rm probe} = K^\dagger \widetilde{H}_{\rm probe} K$, where $K$ is a unitary acting locally
on the qubits and this represents a simplified derivation of a well-known result; see, e.g., the discussion around Eq. (59) in Ref. \onlinecite{Zhang2003}, as well as Refs. \onlinecite{Kraus2001} and \onlinecite{Devitt2006}.

%\textit{Concurrence---}
For quantifying the entanglement in the two-qubit probe state, $\rho$, we use concurrence which is defined as $C(\rho) = \max(0, \lambda_1 - \lambda_2 - \lambda_3 - \lambda_4)$, where $\lambda_1, \lambda_2, \lambda_3, \lambda_4$ are the square roots of the eigenvalues, in decreasing order, of the matrix $R = \rho (\sigma_y \otimes \sigma_y) \rho^* (\sigma_y \otimes \sigma_y)$, with $\rho^*$ being the complex conjugate of $\rho$, and $\sigma_y$ is the Pauli-$y$ matrix. We will study the first and second parametric derivatives of concurrence denoted by $\partial_{g} C$ and  $\partial_{g}^{2} C$. Note these derivatives are evaluated at specific parameter values, $g$ i.e., not generally $g=0$. The second derivative of the concurrence is of particular interest in the following and we define the {\em curvature of entanglement} (CoE) as
\begin{equation}
    \label{CoE1}
    {\rm CoE} \equiv -\partial^2_g C(\rho).
\end{equation}
As mentioned earlier, the negative sign in the definition of CoE ensures that it is positive at the maxima of concurrence.

\section{The relation between QFI and Concurrence for pure states}}

An analytic connection between the QFI and CoE can be established for pure initial states of the two-qubit quantum probe undergoing lossless evolution generated by $H_{\rm probe}$ We consider this case first before exploring the more general case with losses and mixed states of the probe. The eigenstates of the probe Hamiltonian in Eq.~\eqref{HprobeF} are the four Bell states, which we can compactly write as 
\begin{equation}
\label{bell}
|\beta_{ab}\rangle = (|0,b\rangle +(-1)^a|1,\bar{b}\rangle)/\sqrt{2}~~~,
\end{equation}
with $a,b=0,1$ and $\bar{0}=1$,$\bar{1}=0$. 
Note that the computational basis and Bell basis are both defined relative to the rotated basis that makes $\eta$ diagonal. We start with a generic initial state of the probe expressed in the computational and Bell bases respectively as,
\begin{equation}
\label{psi0}
|\Psi^0\rangle = \sum_{m,n=0,1} a_{mn}^0|m, n\rangle = \sum_{a,b=0,1} \beta_{ab} |\beta_{ab}\rangle~~~,
\end{equation}
with $a_{mn}^0 = \langle m,n|\Psi^0\rangle$ and $ \langle \beta_{ab}|\Psi^0\rangle$.  The time evolved state $|\Psi^t\rangle$ is given by 
\begin{equation} 
    \label{psit}
    |\Psi^t\rangle = \sum_{m,n=0,1} a_{mn}^t|m, n\rangle = \sum_{a,b=0,1} \beta_{ab} e^{-ig\omega_{ab}t} |\beta_{ab} \rangle, 
\end{equation}
where
\begin{equation}
    \label{omegas}
    \omega_{ab} = (-1)^a \eta_x - (-1)^{a+b} \eta_y + (-1)^b \eta_z.
\end{equation}
% \begin{align}
%     \omega_{00} &= \eta_x - \eta_y + \eta_z, &
%     \omega_{10} &= -\eta_x + \eta_y + \eta_z, \nonumber\\
%     \omega_{01} &= \eta_x + \eta_y - \eta_z, &
%     \omega_{11} &= -(\eta_x + \eta_y + \eta_z).
% \end{align}
For a pure two-qubit state expressed in the computational basis, the concurrence is given by, 
\begin{equation}
    C(t) = 2|a^t_{00}a^t_{11} - a^t_{01}a^t_{10}|~~~.
\end{equation}
Using Eq.~\eqref{psit} to express $a_{ij}^t$ in terms of $\beta_{ab}$ we obtain,
\begin{equation}
    \label{conc1a}
    C(g,t) = \bigg| \sum_{a,b=0}^1 (-1)^{a+b} \beta_{ab}^2 e^{-2i g \omega_{ab} t}\bigg|
\end{equation}
% Using,
% \[ \partial_x^2 |f(x)| = |\partial_x^2 f(x)| + 2 \delta(f(x))[\partial_x f(x)]^2,\]
% we obtain, 
% \begin{eqnarray}
%     \label{CoEpure1}
%     \partial^2_g C & = & 4t^2 \bigg[  \bigg|\sum_{a,b=0}^1 (-1)^{a+b} \beta_{ab}^2 \omega_{ab}^2 e^{-2i g \omega_{ab} t} \bigg|  \nonumber \\ && \; -\, \delta(C) \bigg( \sum_{a,b=0}^1 (-1)^{a+b} \beta_{ab}^2 \omega_{ab} e^{-2i g \omega_{ab} t}\bigg)^2 \bigg]. \;
% \end{eqnarray}
% Note that the second term in $\partial_g^2C$ is singular since the derivative of $C(g,t)$ is undefined when the concurrence is zero because of the absolute value in Eq.~\eqref{conc1a}. We will not consider these singluar points in the following and so we have 
% \begin{equation}
%     \label{CoEpure2}
%     \partial^2_g C =  4t^2   \bigg|\sum_{a,b=0}^1 (-1)^{a+b} \beta_{ab}^2 \omega_{ab}^2 e^{-2i g \omega_{ab} t} \bigg|, \; C\neq 0.
% \end{equation}

For a pure state the QFI is given by $F(g,t) = 4 t^2 \langle (\Delta h)^2 \rangle$, where $\langle (\Delta h)^2 \rangle$ is the variance of $h$ with respect to $|\Psi^t\rangle$. The factor of $t^2$ appears for the time dependent case because the generator of translations in $g$ is $t \times h$. Using the Eqs.~\eqref{HprobeF} and \eqref{psit}, we obtain
\begin{equation}
    \label{QFIpure1}
    F(g,t) \! =\! 4t^2 \! \bigg[ \sum_{a,b=0}^1 \! |\beta_{ab}|^2 \omega_{ab}^2  - \!\bigg(\! \sum_{a,b=0}^1 |\beta_{ab}|^2 \omega_{ab} \!\bigg)^{\!2} \bigg].
\end{equation}
%In the following subsection we look at the relationship between the CoE and the QFI from Eq.~\eqref{QFIpure1}.

% Commented out text that was here was moved to the end of the document - Anil - 10th Nov

\subsection{Dynamics of Specific Initial States}

\subsubsection{Optimal Probe Dynamics}
Without loss of generality let us assume that $\eta_x \geq \eta_y \geq \eta_z$ with the additional assumption that all three are positive. The first condition can be realized by adding a suitable relabeling of the axes along with the orthogonal transformation that brings $H_{\rm probe}$ to the anisotropic Heisenberg form while the second one is realized by a suitable choice for the zero point for the energy of the system.
%\textcolor{blue}{SKG: can additional words for why we can take eta's to be positive, and assume the ordering here
%be added to justify the without loss of generality remark? Off hand it seems like generality is lost.}
Then from Eq.~\eqref{omegas} we find that the largest and smallest eigenvalues of $\widetilde{H}_{\rm probe}$ are $\omega_{10}$ and $\omega_{11}$ respectively. Since the QFI in Eq.~\eqref{QFIpure1} is proportional to the variance of $\widetilde{H}_{\rm probe}$, we choose an initial state for the quantum probe that is an equal superposition of the eigenstates corresponding to the largest and smallest eigenvalues which is the state that maximizes $\langle (\Delta \widetilde{H}_{\rm probe})^2\rangle$ 
%[REF: Spectral Theorem, Min-Max Theorem for Hermitian Operators]. 
We therefore choose
\begin{equation}
    \label{ProbePure1}
    |\Psi^0_{\rm opt} \rangle = \frac{1}{\sqrt{2}} \big( |\beta_{01}\rangle + |\beta_{11} \rangle \big) = |01\rangle~~~,
\end{equation}
which is simply the initial condition corresponding to one qubit being excited.
%Note that we may introduce a relative phase $\varphi$ between the two components of the state given above but since $\widetilde{H}_{\rm probe}$ is in any case going to generate a relative phase between the two, adding an additional phase in the initial state does not affect any of our subsequent computations and conclusions. 
%\textcolor{blue}{SKG: the above sentence about a relative phase needs to be deleted or clarified since now we have just
%the 01 initial condition.}
Corresponding to this initial state we find the QFI to be
\begin{equation}
F(g,t) = 4t^2 \eta_{xy}^2~~~, 
\end{equation}
where $\eta_{xy} \equiv \eta_x + \eta_y$ and the concurrence as $C(g,t) = |\sin (2g\eta_{xy} t)|$. When $C(g,t) \neq 0$ we can compute the CoE as, 
\begin{eqnarray}
    \label{PureExample1}
    -\partial_g^2 C(g,t) &=& 4t^2\eta_{xy}^2 \sin (2g\eta_{xy}t) {\rm sgn}[\sin (2g\eta_{xy}t)],   \nonumber \\
    & = & 4t^2\eta_{xy}^2 |\sin (2g\eta_{xy}t)|~~~.
\end{eqnarray}
(No $\eta_z$ terms appear in the above expressions because the initial state does not
couple to the corresponding $\sigma_z \otimes \sigma_z$ operator in the Hamiltonian, Eq. (\ref{HprobeF}).)
We see that 
\begin{equation}
\label{inequality}
\rm{CoE}~\leq~ \rm{QFI} ~~~,
\end{equation}
for all values of $t$. 
However it may be noted that when $|\sin (2g\eta_{xy}t)| = 0$ and the concurrence is minimum, the first derivative of concurrence is not defined and the CoE tends to
$+\infty$. 
We further see that the points where the concurrence is maximum, which occurs when  $|\sin (2g\eta_{xy}t)| = 1$, we have the equality with CoE = $F(g,t)$. 

\subsubsection{SLD and Optimal Readout}
%SKG: I'm still worried about our language since there can be cons with SLD eigvecs and they are not ``the'' only ones that can saturate the QCRB. I tried to modify the language a tad to reflect this.
What is the quantum-metrological significance of the points in time when ${\rm CoE}=F(t)$? The answer to this question is obtained by computing the concurrence of the eigenvectors of the SLD which furnish 
an
optimal read-out strategy for the quantum probe 
%in any practical 
quantum metrology schemes designed to saturate the QCRB. The SLD operator expressed in terms of the eigenstates and eigenvalues of the state of the probe density matrix, $\rho = \sum_a \lambda_a |\lambda_a\rangle \langle \lambda_a|$, is 
\begin{equation}
    \label{SLDop}
     L_\rho(\partial_g \rho) = \sum_{a,b} \frac{2 \langle \lambda_a |\partial_g \rho |\lambda_b \rangle}{\lambda_a + \lambda_b} |\lambda_a \rangle \langle \lambda_b|, 
\end{equation}
with the sum again restricted to those terms for which $\lambda_a + \lambda_b \neq 0$. For the particular case where the probe state is pure, this reduces to $L_{\Psi}=2[(\partial_g|\Psi^t\rangle) \langle \Psi^t| + |\psi^t\rangle (\partial_g \langle \Psi^t|)]$ which can be computed using $|\Psi^t_{\rm opt} \rangle = (e^{-ig\omega_{01} t}|\beta_{01}\rangle + e^{-ig\omega_{11} t}|\beta_{11}\rangle)/\sqrt{2}$ to be,
\begin{equation}
\label{SLDexample1}
L(\Psi_{\rm opt}) =2t \eta_{xy} \left(
\begin{array}{cccc}
 0 & 0 & 0 & 0 \\
 0 & - \sin (2 g t\eta_{xy}) &  i \cos (2 g t\eta_{xy}) & 0 \\
 0 & - i  \cos (2 g t \eta_{xy}) &  \sin (2 g t \eta_{xy}) & 0 \\
 0 & 0 & 0 & 0 \\
\end{array}
\right).
\end{equation}
The SLD in the equation above is expressed in the computational basis and we see that it becomes diagonal and its eigenvectors become the computational basis states $|00\rangle$, $|01\rangle$, $|10\rangle$ and $|11\rangle$ when $\sin (2 g t \eta_{xy}) = 1$, which are also the points in time where $\rm{CoE}$ = QFI. This means that the QCRB can be saturated using simple computational basis measurements at these special points. From the point of view of implementing quantum limited metrology, a simple read-out that can saturate the QCRB is significant since, in general, the read-outs of the quantum probe that saturates the bound are measurements on to entangled states of the probe that are difficult to implement in the lab.

%SKG cleaned up the notation for rho in this paragraph relative to before.
 Note that the eigenstates of the SLD operators are not necessarily the only possible choice of readout of the quantum probe that saturates the QCRB. For quantum states that are not of full rank as in the pure-state example considered here, the SLD is unique only in the support of the density matrix $\rho^t_{\rm opt} = |\Psi^t_{\rm opt}\rangle \langle \Psi^t_{\rm opt}|$. However the non-uniqueness of the SLD eigenstates in the null-space (kernel) of $\rho^t_{\rm opt}$ does not concern us here since the part of the readout that projects on to the kernel of $\rho^t_{\rm opt}$ is not of any relevance for the design of any practical metrology protocols since those do not contribute to the measurement statistics. Hence we can as well assume that arbitrary product measurements suffice within the null-space. Even within the support of  $\rho^t_{\rm opt}$ one may be able to choose multiple probe readout strategies that saturate the QCRB. in fact, for this particular example we have considered, projectors on to the computational basis states $|01\rangle$ and $|10\rangle$  $\rho^t_{\rm opt}$ would suffice for all values of $g$ and $t$.
 That is to say, the classical Fisher information formed from these computational basis states
 is easily seen to be equal to the quantum Fisher information.
 % SKG: I don't believe it's true that |01>,|10> are biased as we claim below and also they are really simple to implement which is much in their favour. If I'm wrong we need to give a proof and also explain why unbias vs bias is better, etc.
 % The other claims -- optimal wrt single shot, mulit-copy, multi-parameter -- must be backed up with references that clearly prove them.  I added below a more gentle argument for the SLD which can be deleted if we can point to clear proofs of SLD superiority.
 %\textcolor{purple}{However, the corresponding estimator for $g$ can be shown to be biased while the estimator corresponding to the SLD eigenstate measurements will be unbiased. It can also be shown that the SLD eigenstates are the best measurements for saturating the QCRB if one demands that the information gain about $g$ be optimal irrespective of whether the quantum metrology protocol involves single shot, multi-copy or for that matter, multi-parameter estimation. For these reasons we are particularly interested in the cases where the SLD eigenstates (within the support of $\rho^t_{\rm opt}$ are product states and leads to a simple readout protocol. }
 While projectors formed from the SLD eigenvectors are therefore not necessarily the only projectors
 that can saturate the QCRB, there may be several reasons, depending on the application, when they would be
 preferred.  For example, there is a clear (albeit possibly involved) recipe for their construction, they have a geometric interpretation, and they could provide better convergence properties for finite resources.

By writing $|\Psi^t_{\rm opt}\rangle = (|\beta_{01}\rangle + e^{-2ig\eta_{xy}t}|\beta_{11}\rangle)/\sqrt{2}$, apart from an overall phase, we see that the time evolution generated by $\widetilde{H}_{\rm probe}$ corresponds to a periodic oscillation between the states $|01\rangle = (|\beta_{01}\rangle + |\beta_{11}\rangle)/\sqrt{2}$ and $|10\rangle = (|\beta_{01}\rangle - |\beta_{11}\rangle)/ \sqrt{2}$ with a frequency equal to $2g\eta_{xy}$. The dynamics is confined to the two-dimensional subspace of the four-dimensional Hilbert space spanned by these two states and for this choice of initial state the generic probe Hamiltonian is equivalent to a 
%Jaynes-Cummings type or 
flip-flop Hamiltonian, 
\begin{equation}
    \label{flipflop}
    \widetilde{H}_{\rm probe} = g\eta_{xy}(\sigma_+ \otimes \sigma_- + \sigma_- \otimes \sigma_+).
\end{equation} 
This is a general feature for any initial state $|\Psi_+\rangle = (|\beta_{ab}\rangle + |\beta_{a'b'}\rangle)/\sqrt{2}$ created as an equal superposition of any two Bell states. The time evolution of this initial state will be periodic oscillations between $|\Psi_+\rangle$ and $|\Psi_-\rangle = (|\beta_{ab}\rangle - |\beta_{a'b'}\rangle)/\sqrt{2}$ with a frequency equal to $g(\omega_{ab} - \omega_{a'b'}) \equiv g \eta$ generated by an effective 
flip-flop Hamiltonian of the form $\widetilde{H}_{\rm probe} = g\eta (|\Psi_+\rangle \langle \Psi_-| + |\Psi_-\rangle \langle \Psi_+|)/2$.
%(In this case the two Bell states that are proportionate to computational basis states $|0,0\rangle \pm |1,1\rangle$, which are $|\beta_{00}\rangle$ and $|\beta_{10}\rangle$ with our convention, are degenerate with eigenvalue 0 and so these states do not have any $g$-dependent time evolution.)

% Let's first consider the case where no restrictions are put on the choice of Hamiltonian. For the QFI to serve as the envelope of $|$CoE$|$, one possible class of initial states (see Appendix B) is \(|\psi(0)\rangle = (|\phi_i \rangle + e^{i \varphi}|\phi_j \rangle) /\sqrt{2} \), where \(|\phi_{i,j} \rangle\) are any two of the Bell states in the rotated frame, with non-degenerate eigenvalues \(\lambda_{i,j}\). In such cases, the effective Hamiltonian acts as a flip-flop one in the subspace spanned by \(|\pm\rangle = (|\phi_i \rangle \pm e^{i \varphi}|\phi_j \rangle) /\sqrt{2} \), i.e.
% \begin{equation}\label{flipflop}
%     H_{\rm eff} = g\lambda_+ I + g\lambda_- \left[ |+ \rangle\langle - | + |- \rangle\langle + |\right],
% \end{equation}
% where \(\lambda_\pm=(\lambda_i \pm \lambda_j)/2\). The \(g\lambda_+ I \) terms only adds a trivial global phase and can be ignored. 

For any such flip-flop probe Hamiltonian and initial state $|\Psi_+\rangle$ (or $|\Psi_-\rangle)$, the loss-free dynamics of  entanglement between the two qubits as measured by the concurrence is, $C = \left| \sin(2g\eta t )\right|$, and the corresponding QFI is given by $F = 4\eta^2 t^2$, and for all these cases the inequality Eq. (\ref{inequality}) holds.  We again have equality at those points in time when the concurrence is maximum and $\sin(2g\eta t ) = 1$.

\subsubsection{General Initial States}
\begin{figure}[!htb]
    \centering
    \includegraphics[width=1\linewidth]{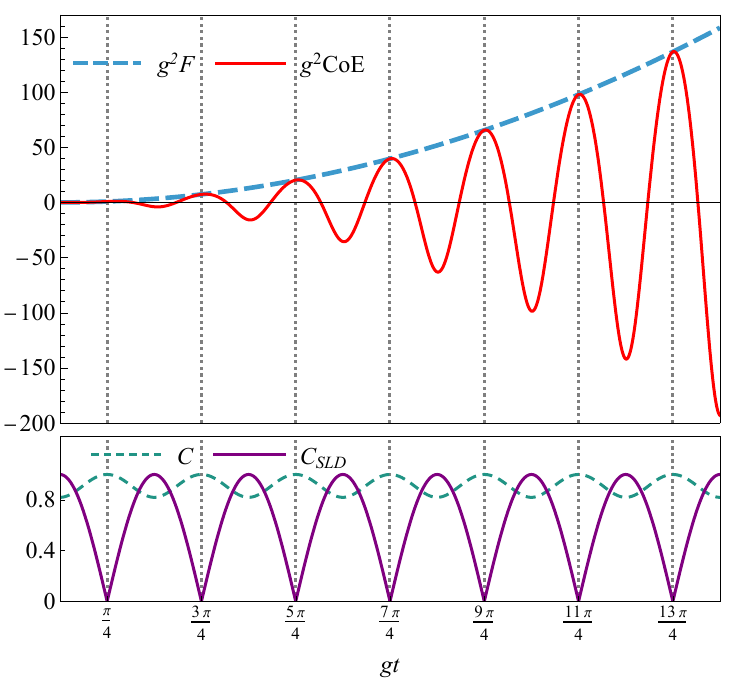}
    \caption{(Color online) Entanglement and QFI dynamics for initial state with unequal superposition of Bell states and $\alpha = 0.3$ with $\eta_{xy}=1$. The upper panel shows the QFI expressed as the unitless $g^2F$ (blue dashed line) which acts as the upper bound for the CoE, also in unitless form $g^2$CoE (red solid line). The concurrence, $C$ (blue dashed line) and the identical concurrences of the two entangled eigenstates of the SLD operators, $C_{\rm SLD}$ (purple line) are shown in the lower panel. The vertical dotted lines serve as a guide to the eye showing that the points where CoE = QFI coincides with the maxima of $C$ and the zeros of $C_{\rm SLD}$ at the points where $\sin(2gt\eta_{xy}) = 1$.}
    \label{UnequalBell1}
\end{figure}

Instead of an equal superposition of two Bell states, we can choose as the initial state an unequal superposition of a pair of Bell states, 
\begin{equation}
 \label{bellsup1}
 |\Psi_{\alpha}^0\rangle = \alpha|\beta_{01}\rangle + \sqrt{1-\alpha^2} |\beta_{11}\rangle~~. 
 \end{equation}
The parameter-dependent evolution of the probe is again described by the Hamiltonian in Eq.~\eqref{flipflop}. For this initial state we find, 
\begin{eqnarray*}
    F(t) & = & 16 \alpha ^2 (1-\alpha ^2) t^2 \eta _{xy}^2, \\
    C(g,t) & = &  \sqrt{1-4 \alpha ^2 \left(1-\alpha ^2\right) \cos ^2 (2 g t \eta_{xy})}.
\end{eqnarray*}
The QFI in this case is maximum when $\alpha = 1/\sqrt{2}$, indicating that the choice of initial state in Eq.~\eqref{ProbePure1} is optimal. The key difference when $\alpha \neq 1/\sqrt{2}$ is that the concurrence is always positive and does not touch zero and its first derivative is  continuous which means that the CoE is finite and well defined at all points in time. The SLD operator in this case turns out to be $2\alpha \sqrt{1-\alpha^2} L_\psi$ where $L_\psi$ is given in Eq.~\eqref{SLDexample1}. This means that the SLD operator is independent of $\alpha$ for the family of initial states, $|\Psi_\alpha^0\rangle$. As before, the SLD operator has the product states $|00\rangle$ and $|11\rangle$ as two of its eigenstates while the other two are entangled states which become the computational basis states $|01\rangle$ and $|10\rangle$ at the points where CoE = QFI. Both these eigenstates have identical values of concurrence as a function of time, which we denote as $C_{\rm SLD}(g,t)$  The QFI, CoE, concurrence as well as $C_{\rm SLD}$ as a function of time is shown in Fig.~\ref{UnequalBell1} for the case $\alpha$ = 0.3 in the initial condition of Eq. (\ref{bellsup1}). In all the figures, we work in terms of the scaled, unit-less variable $gt$ and we use the dimensionless products $g^2 F(g,t)$ and $g^2 \rm{CoE}$ \cite{saleem2022optimal,saleem2024achieving}. These units are
appropriate for the types of Hamiltonian under consideration because by specifying any particular $g$ and $t$ one can
use the figures to construct the corresponding QFI and CoE. From a practical standpoint, one can use one value of
$g$, say $g$ = 1, to construct the figures (see Appendix A of Ref. \onlinecite{saleem2024achieving}).

As a function of $t$, the points where CoE = QFI do not correspond to maxima of the CoE. These maxima may be located by setting $-\partial_t [\partial^2_gC(g,t)]=0$ which, using Eq.~\eqref{PureExample1}, leads to a transcendental equation, $\tan (2g \eta_{xy} t) = -g \eta_{xy} t$. For large values of $t$ the transcendental equation is satisfied when  $gt \sim (2n+1)\pi/4$ for integer values of $n \gg 1$. At these points in time the absolute value of CoE touches the QFI very close to the maxima, as a function of time, of the CoE. This can be seen from Fig.~\ref{UnequalBell1} wherein for small values of $gt$ the points where CoE touches QFI are slightly to the left of the peaks of CoE while as $gt$ becomes larger the point of contact moves progressively closer to the corresponding maxima of CoE.

\subsubsection{Coincidence Properties}
 In summary, for the time-evolving two-qubit systems with initial conditions considered above, there are times $t$ when the following four properties
 are satisfied simultaneously:
 \begin{enumerate}
    \item The concurrence, when viewed as a function of $g$, is a maximum.
    \item The CoE with respect to $g$ is a maximum.
    \item CoE = QFI or $-\partial^2_g C(g,t)  =   F(t)$.
    \item The concurrence of all SLD eigenstates, $C_{\rm SLD}$, is zero.
 \end{enumerate}
 We refer to these four properties collectively as {\em coincidence properties}.

\begin{figure}[!htb]
    \centering
    \includegraphics[width=1\linewidth]{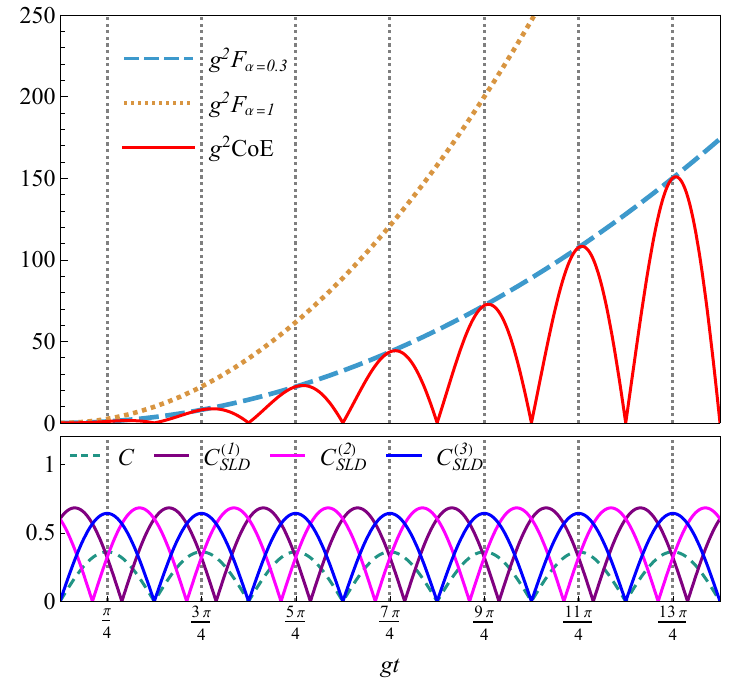}
    \caption{Entanglement and QFI dynamics for the initial state $|\Phi_{\alpha}^0\rangle$ with $\alpha = 0.3$ and $\eta=1$. The upper panel shows the QFI expressed as the unitless $g^2F_{\alpha =0.3}$ (blue dashed line) which acts as the upper bound for the CoE, also expressed in unitless form $g^2$CoE (red solid line). For comparison, also shown is the QFI corresponding to the the optimal initial state, $g^2F_{\alpha =1}$ (gold dotted line). The concurrence, $C$ (blue dashed line) and the three distinct values of concurrences of the three entangled eigenstates of the SLD operators, $C_{\rm SLD}^{(1)}$ (purple line), $C_{\rm SLD}^{(2)}$ (magenta line) and $C_{\rm SLD}^{(3)}$ (blue line) are shown in the lower panel. The vertical dotted lines serve as a guide to the eye showing that the points where CoE = QFI coincides with the maxima of $C$ when $\sin(2gt\eta_{xy}) = 1$. We see that at these points the SLD eigenstates do not become product states.} 
    \label{GenProdState}
\end{figure}

 All four of the coincidence properties enumerated above need not be satisfied for more general families of states. In particular the fourth property is violated for the class of initial states given by
 \begin{eqnarray}
     \label{phifamily}
     |\Phi^0_\alpha \rangle \!\!& = &\!\! (\alpha |0\rangle + \sqrt{1-\alpha^2}|1\rangle)\otimes |1\rangle \nonumber \\
     &= & \!\! \frac{\alpha}{\sqrt{2}}(|\beta_{01}\rangle + |\beta_{11}\rangle) + \frac{\sqrt{1-\alpha^2}}{\sqrt{2}}(|\beta_{00}\rangle - |\beta_{10}\rangle).  \qquad
 \end{eqnarray}
 We consider evolution of the probe under the flip-flop Hamiltonian from Eq.~\eqref{flipflop} with $\eta_x = \eta_y = \eta/2$ so that $\eta_{xy} = \eta$ and find that $C(g,t) = \alpha^2 |\sin(2 g \eta t)|$ and, 
 \[ F(t) = 4 \alpha^2 t^2 \eta^2, \quad  {\rm CoE} = 4 \alpha^2 t^2 \eta^2 |\sin(2 g \eta t)|, \]
 Except when $\alpha = 1$, these states are not optimal for sensing since the QFI is suppressed by a factor $\alpha^2$ and $|\Phi_1^0\rangle = |\Psi_{\rm opt}^0\rangle$. The key difference from the family of states $|\Psi_{\alpha}\rangle$ considered earlier is that these states never become maximally entangled under time evolution generated by the flip-flop Hamiltonian and $C(g,t) \leq \alpha^2$. This can be attributed to the non-entangled component $\sqrt{1-\alpha^2}|11\rangle$ which does not evolve under the dynamics considered. We also find that SLD operator has as its eigenstates $|00\rangle$ and three entangled states, in general, with distinct values and time dependence for their respective concurrences. The QFI, CoE, concurrence and the three $C_{\rm SLD}$ values are plotted in Fig.~\ref{GenProdState} for the case $\alpha$ = 0.3 in the initial condition of Eq. (\ref{phifamily}). We see that while the first three of the coincidence properties are simultaneously satisfied, as well as the inequality Eq. (\ref{inequality}), the fourth coincidence property is not satisfied in this case. 
 
%Of course we have not exhausted all the possible initial states for which the relationship between QFI and CoE can exist but have provided a procedure to test any given initial condition.  
It should be pointed out that the families of initial states we consider are written in the basis in which $\widetilde{H}_{\rm probe}$ has the form given in Eq. (\ref{HprobeF}). This is not necessarily same as the computational basis defined by the generic two-qubit interaction Hamiltonian given in Eq.~(\ref{inter}). To obtain an initial state in the computational basis defined by Eq. (\ref{inter}) one would have to determine the specific transformation matrix $K$ that converts $H_{\rm probe}$ into the form in Eq.~ (\ref{HprobeF}).  It is important to note, however, that the general anisotropic Heisenberg form of $\widetilde{H}_{\rm probe}$ encompasses many important problems \cite{Cole2006} and so even the case of $K = \openone$ is quite relevant.

%SKG commented this out since IIIC proves the inequality.
%In the above examples, regardless of whether all or any of the coincidence properties are satisfied,  we  Indeed we have examined millions of randomly generated Hamiltonians, i.e., choices of $\eta_x$, $\eta_y$ and $\eta_z$ in Eq. (\ref{HprobeF}), millions of complex-valued random initial conditions, i.e., $\beta_{a,b}$ in Eq. (\ref{psi0}),  and large ranges of $gt$ values. The inequality Eq. (\ref{inequality}) has never been seen to be violated. This extensive scan is made possible by the general pure-state analytical results for the QFI, Eq. (\ref{QFIpure1}), and the concurrence, Eq. (\ref{conc1a}), the latter of which can be differentiated to lead to a somewhat involved but exact analytical form for the CoE. We realize that this does not constitute a proof but does represent strong support for us to conjecture that Eq. (\ref{inequality}) is an actual general bound for the pure state case.

%\vspace{-3 mm}
\vspace{3 mm}

\subsection{General proof of the bound and saturation conditions}
\label{sec:generalbound}

The inequality in Eq.~\eqref{inequality} is not restricted to the specific families of states considered above, but in fact holds for arbitrary pure two-qubit states evolving under Eq.~\eqref{HprobeF}. The key observation is that the same spectral fluctuations that determine the QFI in Eq.~\eqref{QFIpure1} also govern the parametric curvature of the concurrence in Eq.~\eqref{conc1a}. 

To make this connection explicit, it is convenient to rewrite the QFI in terms of the centered frequencies
\begin{equation}
\Delta_{ab} \equiv \omega_{ab}-\mu~~, \qquad 
\mu \equiv \sum_{a,b} |\beta_{ab}|^2 \omega_{ab}~~,
\end{equation}
so that Eq.~\eqref{QFIpure1} becomes
\begin{equation}
F(g,t)=4t^2 \sum_{a,b} |\beta_{ab}|^2 \Delta_{ab}^2~~.
\label{QFIcentered}
\end{equation}
The QFI is therefore proportional to the variance of the interaction spectrum sampled by the initial state.

Next, consider the concurrence amplitude appearing in Eq.~\eqref{conc1a}. Since an overall phase does not affect the concurrence, we may remove the mean frequency $\mu$ from the dynamics by multiplying the complex amplitude by $e^{2igt\mu}$. Define the shifted complex concurrence amplitude
\begin{equation}
\widetilde{C}(g,t)
\equiv
\sum_{a,b}
(-1)^{a+b}\beta_{ab}^2
e^{-2igt\Delta_{ab}}~~,
\label{complexconc}
\end{equation}
so that the physical concurrence is simply
\begin{equation}
C(g,t)=|\widetilde{C}(g,t)|~~.
\end{equation}
The concurrence therefore depends only on the centered spectral fluctuations $\Delta_{ab}$.

Differentiating Eq.~\eqref{complexconc} twice with respect to $g$ produces factors of $\Delta_{ab}^2$, precisely the same quantities that appear in Eq.~\eqref{QFIcentered},
\begin{equation}
\partial_g^2 \widetilde{C}(g,t)
=
-4t^2
\sum_{a,b}
(-1)^{a+b}
\beta_{ab}^2
\Delta_{ab}^2
e^{-2igt\Delta_{ab}}~~.
\label{complexsecond}
\end{equation}
Taking the modulus and applying the triangle inequality immediately yields
\begin{align}
\left|
\partial_g^2 \widetilde{C}(g,t)
\right|
&\leq
4t^2
\sum_{a,b}
|\beta_{ab}|^2
\Delta_{ab}^2
\nonumber\\
&=
F(g,t).
\label{trianglebound}
\end{align}

The remaining step is to relate the curvature of the modulus to the modulus of the second derivative. For any twice differentiable complex function $f(g)$ with $f(g)\neq 0$, one has
\begin{equation}
-
\partial_g^2 |f(g)|
\leq
|f''(g)|~~.
\label{modulusineq}
\end{equation}
Applying Eq.~\eqref{modulusineq} to the concurrence amplitude in Eq.~\eqref{complexconc} gives the general bound
\begin{equation}
{\rm CoE}(g,t)
=
-\partial_g^2 C(g,t)
\leq
F(g,t)~~,
\label{generalbound}
\end{equation}
valid at all points where the concurrence is nonzero and twice differentiable. Points where $C(g,t) = 0$ require separate treatment because the absolute value in Eq.~\eqref{conc1a} generates cusps in the concurrence, as already seen in the example leading to Eq.~\eqref{PureExample1}.

The saturation conditions follow directly from the two inequalities used above. 
First, define
\begin{equation}
X_{ab}(g,t)
\equiv
(-1)^{a+b}\beta_{ab}^2 e^{-2igt\Delta_{ab}}~~ .
\end{equation}
Then Eq.~\eqref{complexsecond} is proportional to
$\sum_{a,b}\Delta_{ab}^2 X_{ab}$. The triangle inequality in Eq.~\eqref{trianglebound} is saturated when all nonzero terms in this sum have the same phase,
\begin{equation}
\arg\!\left[\Delta_{ab}^2 X_{ab}(g,t)\right]
=
\arg\!\left[\Delta_{a'b'}^2 X_{a'b'}(g,t)\right]~~,
\label{sat_triangle}
\end{equation}
for all contributing pairs $(a,b)$ and $(a',b')$. Since $\Delta_{ab}^2$ is real and non-negative, this condition simply states that all spectral components contributing to the concurrence curvature interfere constructively, with no destructive interference between different frequency sectors. For the first pure-state example we considered with the initial state given in Eq.~\eqref{ProbePure1}, this condition corresponds to having $4 \eta_{xy}gt = \pi$, or equivalently $|\sin(2\eta_{xy}gt)|=1$ which are indeed the points where the four coincidence criteria are met.

The second saturation condition comes from Eq.~\eqref{modulusineq}. Writing the concurrence amplitude in polar form,
\begin{equation}
\widetilde{C}(g,t)
=
R(g,t)e^{i\phi(g,t)},
\end{equation}
with $R(g,t)=C(g,t)$, the inequality in Eq.~\eqref{modulusineq} is saturated when the complex amplitude bends only in the radial direction,
\begin{equation}
\partial_g\phi(g,t)=0,
\qquad
\partial_g^2\phi(g,t)=0,
\qquad
\partial_g^2R(g,t)\leq 0 .
\label{sat_radial}
\end{equation}
Equivalently, the phase of $\widetilde{C}(g,t)$ is locally stationary up to second order, so the curvature is entirely associated with changes in the magnitude of the concurrence amplitude rather than rotations in the complex plane. For the first example we considered, the first two conditions are always met since $\widetilde{C}(g,t) = \sin(2gt\eta_{xy})e^{i\frac{\pi}{2}}$ while the last condition is true when $\sin(2gt\eta_{xy}) \geq 0$ including at the maxima of the concurrence amplitude.

For the flip-flop dynamics discussed in Sec.~III~A, these conditions are realized precisely at the points where the concurrence reaches its maximum value. At those times, the entanglement generated between the two qubits is completely coherent, all spectral components interfere constructively, and the entanglement curvature directly reproduces the full metrological susceptibility encoded in the QFI. 

\subsection{CoE, QFI and Bures Distance}

In the above we have found that there are points in time when ${\rm CoE} \, = \, {\rm QFI}$. At these times the concurrence, $C$, viewed as a function of $g$, is a maximum.  Consider one such time, $t$= $t_m$:
\begin{equation}
    \label{key}
    -\partial^2_g C(g,t=t_m)  =    F(g,t=t_m) ~~.
\end{equation}
We consider $t_m$ to be fixed and so drop reference to it for simplicity.
Another meaning to the above relation can be found as follows. Close to $g$, using a Taylor expansion about $g$, we have
\begin{equation}\label{bures2}
    C(g +\delta g) - C(g) = \frac{1}{2} \partial_g^2 C(g)\delta g^2 =  - \frac{1}{2}F \delta g^2~~,
\end{equation}
where $\partial_g C(g) = 0$ at the maximum has been used.
The QFI, $F$, can be related to the Uhlmann fidelity, ${\mathcal F}_U$ as \cite{Braunstein1994},
\begin{equation}\label{bures3}
     F(g) = \dfrac{8}{\delta g^2}\Big[1-\sqrt{{\mathcal F}_U(\rho_{g+\delta g},\rho_g)} \, \Big]~~,
\end{equation}
where 
\[ {\mathcal F}_U(\sigma, \chi) \equiv\Big[\text{Tr} \sqrt{\sqrt{\sigma} \chi \sqrt{\sigma}}\Big]^2~. \]
Substituting into Eq. (\ref{bures2}) and rearranging gives us
\begin{equation}\label{confid}
     C(g)-C(g+\delta g) = 4 \big( 1-\sqrt{{\mathcal F}_U(\rho_{g+\delta g},\rho_g)} \,\big)~~.  
\end{equation}
Introducing the Bures distance~\cite{Braunstein1994}, $d_B(\rho_{g+\delta g},\rho_g)$, a measure of the separation of the density matrices $\rho_{g+\delta g}$ and $\rho_g$,
we have:
    \begin{equation}\label{conbur}
     C(g)-C(g+\delta g) = 2 d_B(\rho_{g+\delta g},\rho_g)^2~~.  
\end{equation}
Thus  loss of entanglement as one moves away from the maximum at $g$ to $g+\delta g$ is equal to twice
the square of the Bures distance between the density matrices when Eq. (\ref{key}) holds, an appealing geometric interpretation.  It is also easy to start with Eq. (\ref{conbur}) (or Eq. (\ref{confid})) and work backwards to
arrive at Eq. (\ref{key}); i.e., one can view
these two relations as being equivalent. That is, Eq. (\ref{key}) {\em iff} Eq. (\ref{conbur}).

Notice that the considerations in this subsection are quite general, i.e., they are not constrained to pure state systems as we considered in the previous sections. Furthermore there is no explicit requirement on the two-qubit Hamiltonian.

\vspace{0.5cm}

\section{Open evolution of the quantum probe}

Let us consider the flip-flop Hamitlonian give in Eq \ref{flipflop} with $\eta_{xy}=1$, and consider the open system case. In addition to the parameter-dependent time evolution generated by Hamiltonian on the quantum probe, we assume that the probe is also subject to Markovian noise described by the Gorini-Kosskowski-Sudarshan-Lindblad (GKSL) equation \cite{gorini_completely_1976, lindblad_generators_1976, chruscinski_brief_2017, manzano2020short},
\begin{equation}
\dot{\rho} = {\mathcal L}\rho= -i [ H, \rho ] +  \kappa \sum_{j=1}^2 \D\left[\sigma_-^{(j)}\right](\rho),
\label{Lindblad2}
\end{equation}
where ${\mathcal L}$ is the generator of the open evolution.  The dissipator, ${\mathcal D}$, is given by  $\D[\O](\rho) = \O \rho \O^\dag - (\O^\dag \O \rho + \rho \O^\dag \O)/2$, while $\kappa$ represents the decay rate of the qubits to their respective ground states. Note that we have assumed the same decay rate for both the qubits.  The pure state analysis of Secs. II is
not directly applicable.  However, the evolution equations are still linear and, for the cases considered of sufficiently
low dimension that analytical solutions can be found and simplified with symbolic manipulation tools; we used Mathematica \cite{Mathematica}.
%Cases where different decay rates are assumed for each of the qubits are included in the 
%End Matter.
%Appendix B.

\subsection{Separable Initial State} 

Let us consider the separable and optimal state give in Eq. \ref{ProbePure1} as the initial state of the probe, that is, $\rho_{\rm opt}(0) = |01\rangle \langle 01|$. Explicit exponentiation of the generator ${\mathcal L}$ describing the open evolution is possible and the time evolved state $\rho_{\rm opt}(t) = e^{{\mathcal L}t} \rho_{\rm opt}(0)$ of the probe can be computed. The non-zero elements of $\rho_{\rm opt}(t)$ are:
\begin{eqnarray*}
	\rho_{\rm opt}^{11}(t) & = & 1-e^{-\kappa t} \\ 
	\rho_{\rm opt}^{22}(t) & = &   e^{-\kappa t} \sin^2(gt) \\
	\rho_{\rm opt}^{33} (t) & = &   e^{-\kappa t} \cos^2(gt) \\
%	\rho_{\rm opt}^{12}= \rho_{\rm opt}^{21*} & = & i  \sin(gt) e^{-\frac{\kappa t}{2}}  \\
%	\rho_{\rm opt}^{13}= \rho_{\rm opt}^{31*} & = &  \cos(gt) e^{-\frac{\kappa t}{2}} \\
	\rho_{\rm opt}^{23}= \rho_{\rm opt}^{32*} & = &  -i e^{-\kappa t} \sin(2gt)/2.
\end{eqnarray*}
Using $\rho_{\rm opt}(t)$ we  calculate the QFI as a function of time and obtain,  
\begin{equation}
    F(t) = 4  t^2  e^{ - \kappa t} 
\end{equation}
We can also calculate the concurrence of this state as, 
\begin{equation}
  C(g,t)=  e^{- \kappa t}  \big|\sin(2 gt)\big|,
\end{equation}
and except at the points where $\sin(2gt)=0$,
\begin{equation} 
{\rm CoE}  =  F(t)  |\sin(2 g t)|.
\end{equation}
In the noisy case also we again see that ${\rm CoE} \, = \, {\rm QFI}$ at the maxima of the concurrence.

% We see a direct connection between the curvature of entanglement and QFI. 
% At the points where $G_s(gt) = 1$, we have CoE equal to $-F_s(t)$. These points correspond to $gt = (2n+1)\pi/4$ for $n=0,1,\ldots$. At these points the CoE itself have minima as a function of $g$ and from Eq.~\eqref{firstder} we see that at these points $\partial_g C_s(t) = 0$ and so the concurrence has maxima as a function of $g$ when $gt = (2n+1)\pi/4$. 

% As a function of $t$, the points where $\partial^2_gC_s(g,t)= -F_s(t)$ do not correspond to minima of the CoE. These minima may be located by setting 

 In this case, setting $\partial_t [\partial^2_gC_s(g,t)]=0$  leads to the transcendental equation, $\tan (2gt) = 2gt/(\kappa t-2)$. The right hand side of this equation is large in magnitude except for very small values of $t$ provided $g/\kappa$ is large. This is due to the function having a simple pole at $\kappa t = 2$ and it tending to $g/\kappa$ for large values of time. This again means that except for small values of $t$, the transcendental equation is satisfied when  $gt \sim (2n+1)\pi/4$ such that the points in time where the absolute value of CoE touches the QFI are very close to the minima of CoE. When $g/\kappa$ is large we find that the the maxima of the concurrence also aligns closely with the points where CoE is equal to $-F_s(t)$.

The SLD operator in this case is equal to $L(\Psi_{\rm opt})$ from Eq.~\eqref{SLDop} and two of its eigenvectors are the product states $|00\rangle$ and $|11\rangle$. The remaining two eigenstates are typically entangled and they have identical concurrences given by,
\begin{equation}
    C_{\rm SLD}(t) = |\cos(2 g t)|.
\end{equation}

\begin{figure}[!htb]
    \centering
    \includegraphics[width=\linewidth]{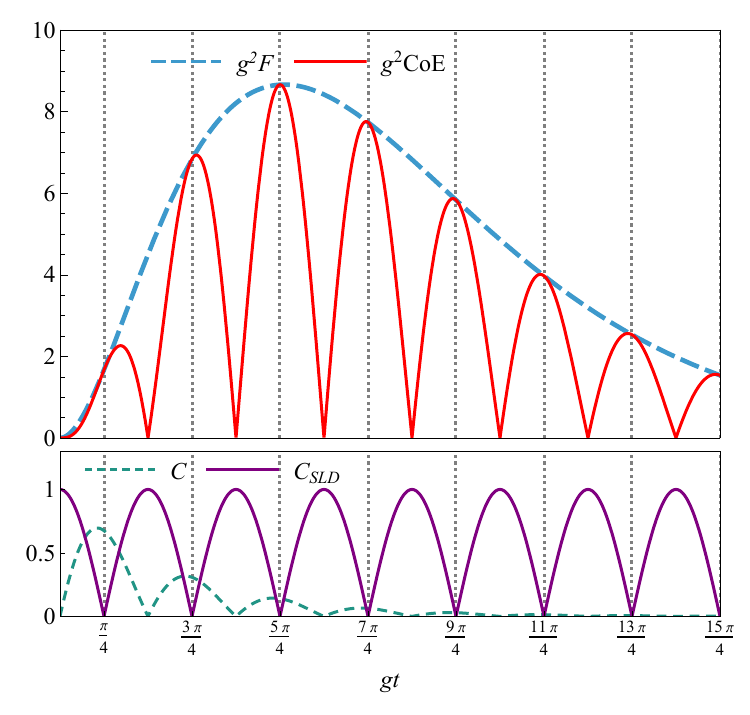}
    \caption{Entanglement and QFI dynamics for an initially separable state subject to dissipation. Upper panel: 
    the QFI, expressed in unitless form $g^2F$ (blue dashed line), serves as the envelope for CoE, also in unitless form $g^2$CoE (red solid line). The CoE is not defined at the points where $\sin(2gt)=0$. Lower panel: concurrence of the system state $\rho_s(t)$ (green dashed line) and the non-zero concurrence of two of the eigenvectors of SLD (purple solid line).  Vertical dotted lines denotes points where CoE = QFI and $C_{\rm SLD}$ is zero. The decay rate is set to $\kappa /g =0.5$ in both panels.}
    \label{QFIsep}
\end{figure}

In Fig.~\ref{QFIsep}, we plot the QFI, the CoE, the concurrence of the probe state and $C_{\rm SLD}(t)$ as a function of the scaled time $gt$. We also use scaled values of the loss parameter as $\kappa g$.  We see that the CoE is enveloped by the QFI except at the points where the CoE is singular. These singularities being point-like are however not seen in the plots. We also see that the points where CoE = QFI line up quite well with the maxima of the concurrence except for small $t$ when the point of contact is slightly to the right of the concurrence maxima.  We again see that when  CoE = QFI, $C_{\rm SLD}(t) =0$, indicating that simple, product-state measurements suffice to saturate the QCRB at these instants in time in an implementation of a quantum metrology scheme to estimate $g$ using the two-qubit probe even in the presence of decoherence. All four of the coincidence properties are simultaneously satisfied in this case. It may be noted that even if the concurrence does not become equal to 1 at any time, it does become equal to the maximum possible value as allowed by the open evolution at the points where all the coincidence properties are satisfied. Previously~\cite{saleem2022optimal, saleem2024achieving} it was noted that with losses it is advantageous to choose the point in time where the QFI is a maximum as the optimal time to do the probe readout. Here we show that choosing the time point which satisfies the relation CoE = QFI and lies closest to the maximum of $F(t)$ can furnish the dual advantages of near-optimal sensitivity as well as easy to implement probe readouts. 

\subsection{Entangled Initial State}
\begin{figure}[!htb]
    \centering
    \includegraphics[width=\linewidth]{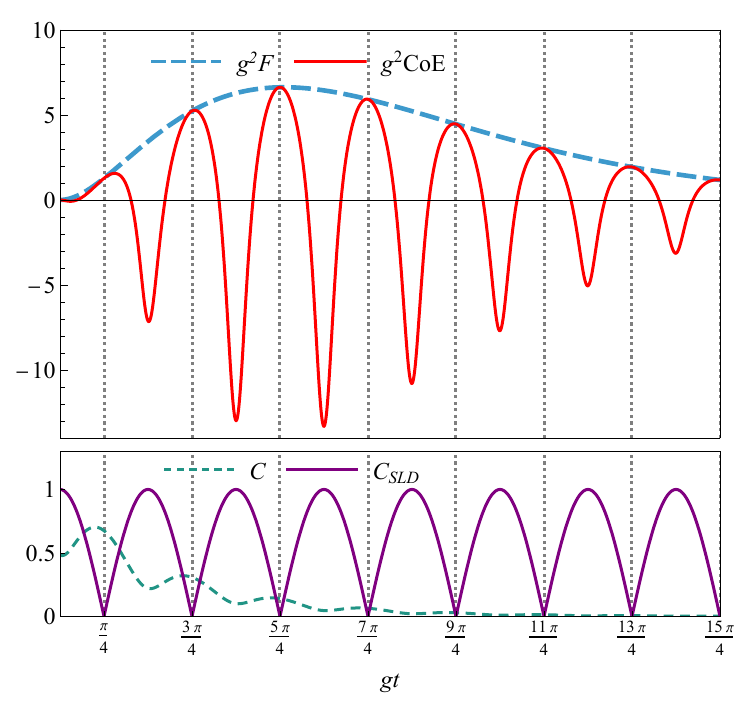}
    \caption{Entanglement and QFI dynamics for an initially entangled state. Upper panel:  The CoE (red solid line) touches the QFI (blue dashed line) at the maxima of the concurrence $C(g,t)$ shown in the lower panel (green dashed line). The lower panel also shows the concurrence of the eigenvectors of the SLD corresponding to non-zero eigenvalues  (purple solid line).  Vertical dotted lines denotes points where CoE = QFI and the four coincidence properties are satisfied. The decay rate is set to $\kappa /g =0.5$ and $\alpha = 0.25$ in both panels.}
    \label{QFIent}
\end{figure}

We consider the initially entangled state given by $\rho_{e,\alpha}(0) = |\Psi_{e,\alpha}(0)\rangle \langle \Psi_{e,\alpha}(0)|$, where,
\begin{equation}
\label{bellsup2}
  |\Psi_{e,\alpha}(0) \rangle = \alpha |01\rangle +  \sqrt{1 - \alpha^2}|10\rangle 
\end{equation}
This initial state 
%can also be expanded in terms of a superposition of bell basis defined in Eq \ref{bell}.
is also an unequal superposition of two Bell states analogous to Eq. \ref{bellsup1} although with
a trivially different choice for $\alpha$.
The time evolution of the density matrix $\rho_{e,\alpha}(t)$ is described by the GKSL master equation in Eq.~\eqref{Lindblad2} and it has the following non-zero elements: 
\begin{eqnarray*}
	\rho_{e,\alpha}^{11}(t) & = & 1 - e^{-\kappa t}\\ 
	\rho_{e,\alpha}^{22}(t) & = &   \frac{e^{-\kappa t}}{2}\Big[1-(1-2\alpha^2)\cos(2gt)\Big]\\
	\rho_{e,\alpha}^{33} (t) & = &   \frac{e^{-\kappa t}}{2}\Big[1+(1-2\alpha^2)\cos(2gt)\Big] \\
	\rho_{e,\alpha}^{23}= \rho_{e,\alpha}^{32*}\!\! & = &\!\!  \frac{e^{-\kappa t}}{2}\Big\{2\alpha\sqrt{1-\alpha^2} +  i\sin(2gt) \big[ 2\alpha^2 -1 \big] \Big\}~~~,
\end{eqnarray*}
where the density matrix indices 1, 2, 3, 4 refer to the usual computational basis components of
$|00\rangle$, $|01\rangle$, $|10\rangle$ and $|11\rangle$.
We can use this density matrix to calculate the QFI,
\begin{equation}
    F(t) = 4 t^2  e^{ - \kappa t} ( 1 - 2\alpha^2 )^2
\end{equation}
The concurrence and CoE are respectively,
\begin{eqnarray}
    C_e(g,t)& = & e^{-\kappa t} \sqrt{1-(1-2 \alpha ^2)^2 \cos^2 (2 g t)}, \nonumber \\ 
    -\partial^2_g C_e(g,t) & = &  F(t)G(gt) ,
\end{eqnarray}
where 
\begin{equation}
    G(gt) = \frac{1-2\cos^2(2gt)+(1-2\alpha^2)^2\cos^4(2gt)}{\big[1-\left(1-2 \alpha ^2\right)^2 \cos^2 (2 g t) \big]^{3/2}}
\end{equation}
We again see that when $G(gt)=1$ at $gt = (2n+1)\pi/4$ ($\cos(2gt)=0$), the CoE is equal to QFI. As functions of $g$, the concurrence $C(g,t)$ has maxima and the CoE itself has minima at these points. As in the case of the previous examples the alignment of the points where CoE = QFI with the maxima of $C(g,t)$ and the minima of CoE itself as functions of time is approximate with good alignment of the three points except for small values of $t$. Note that the CoE is not singular at the minima of the concurrence which corresponds to $\cos(2gt) = 1$.  The expression for the concurrence of the eigenstates of SLD, $C_{\rm SLD}$ turns out to be exactly the same as in the previous case. This means that even when the initial state is entangled, the points in time where CoE is equal to QFI are again operationally significant and correspond to cases where the QCRB can be saturated using product measurements on the two-qubit quantum probe. In Fig.~\ref{QFIent} we plot the CoE, $F(t)$, $C(g,t)$ and $C_{\rm SLD}$ for $\alpha = 0.25$. We see from the figure that all four coincidence properties are satisfied in this case also. It may be noted that in this case also, if the decay is zero, the concurrence can reach the maximum value of 1. 

For the case with decay also there are more general examples we can consider but in all cases we conjecture that CoE $\leq$ QFI with equality at specific points. For open dynamics, in both the examples considered we had all four of the coincidence properties satisfied simultaneously. The fact that in both cases the concurrence will achieve the maximum value possible of 1 in the absence of decay appears to facilitate this. If the initial state is chosen such that the concurrence is less than 1 even in the absence of decay, then as with the pure state example we considered, the fourth coincidence property is not, in general satisfied. It is worth pointing out that we have considered only one type of decay in the two examples, namely the amplitude damping generated by $\sigma_-$ operator with identical decay rates on both qubits. It is worth emphasizing here that the $\sigma_-$ operators are defined in the same basis in which $\widetilde{H}_{\rm probe}$ has the anisotropic Heisenberg form (see Appendix B). For other types of decay channels also all four coincidence properties need not be simultaneously satisfied.

% \purple{It is worth pointing out that the analysis carried out for the general Hamiltonian for the closed system in the previous section can still apply, albeit with corresponding a rotated system-bath coupling operator with the same rotation as Eq.~\eqref{rotate_h}. We gave a concrete example in Appendix~\ref{sec_appx_openr}.}

%\section{Conclusions}\label{sec:conclusions}
\section{Conclusions}
In this work, we looked at a minimal model in which the parameter associated with a many body term of the probe Hamiltonian is estimated. Specifically, we have investigated the relationship between dynamically generated entanglement, as opposed to the entanglement in the initial probe state, and QFI in the simple setting of two qubits interacting through time- and space-independent couplings. Our analysis of entanglement dynamics is in terms of the curvature of entanglement (CoE), defined as the negative of the second derivative of concurrence with respect to the coupling parameter.  We uncovered several notable connections between CoE and QFI for for generic two-qubit Hamiltonians and different families of initial states, including examples in the presence of amplitude damping.  In particular, for these cases, we find that at points in time when concurrence is a maximum when viewed as a function of the coupling parameter, CoE = QFI. We also proved that for the two-qubit, pure-state case, for general nonlocal interaction Hamiltonians and all possible initial conditions, CoE $\leq$ QFI. 
%Based on extensive numerical evidence we conjecture that this relationship holds universally for two-qubit probe states}. 

By examining the time evolution of the concurrence of the eigenstates of the symmetric logarithmic derivative (SLD), we showed that for several families of initial probe states, simple product measurements are not only sufficient but optimal as well for saturating the quantum Cram\'er-Rao bound at the special times when CoE = QFI and otherwise entangled measurements may be required. 
% Again, SKG is trying to be a little cautious with a rephrasing of the below sentence
%This observation has  has practical significance since it can simplify the design of quantum limited metrology schemes that aim to saturate the QCRB optimally in single- or multi-shot estimation protocols.
This observation could be of practical significance when it is desirable to use the SLD
eigenvectors as the projective measurements.

Future work will explore the relation of QFI and entanglement for systems with $N$ $>$ 2 qubits \cite{guhne2009entanglement,hauke2015measuring}, parameters other than the coupling, as well
as the implications of different types of loss. 

Our findings suggest a deeper, time-resolved interplay between the dynamically generated entanglement and measurement precision when the probe Hamiltonian is not restricted to being local, offering a new operational perspective on the role of entanglement in quantum metrology protocols designed to estimate multi-party couplings.

% SKG: The discussion below seemed overly technical and confusing to me and it might be best just to finish with the strong "Our findings suggest ... statement.
%\textcolor{purple}{In the condensed matter context, the QFI estimated from measurable quantities like dynamical susceptibility is limited to being only a witness of multipartite entanglement in the initial probe state since the QFI that is estimated corresponds to a suitably chosen local Hamiltonian and it does not correspond to the native Hamiltonian of the system which has many-body couplings in it.This is primarily because the estimate of the QFI corresponding to a local Hamiltonian involves measurements of two-point correlation functions like susceptibilities and the QFI corresponding to a two-body coupling would require measurements of four-point correlation functions which are beyond the capability of current spectroscopic methods. However, our work on the minimal model indicates that at the points in time when CoE = QFI it may be possible to extract the QFI corresponding to a two-body coupling from the measurement of two-point correlations. This is another interesting line of development to be pursued in the future.}

\begin{acknowledgments}
\vspace{-2.5 mm}
This material is based upon work supported by the U.S. Department of Energy Office of Science National Quantum Information Science Research Centers.  
Work performed at the Center for Nanoscale Materials, a U.S. Department of Energy Office of Science User Facility, was supported by the U.S. DOE Office of Basic Energy Sciences, under Contract No. DE-AC02-06CH11357. A. S. was supported by a grant from the national quantum mission of the Department of Science and Technology, Government of India. D.L. and T.Y were supported in part by ACC-New Jersey under Contract No. W15QKN-18-D-0040.
\end{acknowledgments}

%\vspace{2 cm}

\appendix
\section{Local \texorpdfstring{$SU(2)$}{SU(2)} rotations following the SVD decomposition} \label{sec_appx_su2}
% \vspace{-2 mm}
Here we show that the \(SO(3)\) rotation of the coefficient matrix \(\eta\) in Eq.~\eqref{inter} translates to local \(SU(2)\) transforms on each qubit. Following the \(SO(3)\) transform in Eq.~\eqref{rotate_h}, the rotated spin operators follow the commutation relationship (e.g. for qubit $1$)
\begin{align}
    \left[\tilde\sigma_{j}, \tilde\sigma_{k}\right] &= \left[U_{ji}^T\sigma_{i}, U_{km}^T\sigma_{m}\right] \\
    &= U_{ji}^T U_{km}^T 2i \epsilon_{iml} \sigma_l = 2i \epsilon_{iml} U_{ij} U_{mk} \sigma_l \\
    &= 2i \epsilon_{jkp} U^T_{pl} \sigma_l = 2i \epsilon_{jkp} \tilde\sigma_{p}
\end{align}
where \(\epsilon_{ijk}\) is the Levi-Civita symbol and we have used \(\det(U)=1\), \(\left[\sigma_j, \sigma_k\right] = 2i \epsilon_{jkl} \sigma_l\) and
\begin{align}
    \epsilon_{ijk}\det(U) &= \epsilon_{lmn}U_{li} U_{mj} U_{nk} \\
    \Rightarrow \epsilon_{ijk} U^T_{kn} &= \epsilon_{lmn} U_{li} U_{mj} U_{nk} U^T_{kn} \\
    &= \epsilon_{lmn} U_{li} U_{mj}
\end{align}
Therefore, the \(SO(3)\) rotations on the coefficients can be rewritten as local unitary transformations on the operator basis of the qubits, and the modified spin operators still follow the same algebra of the Pauli matrices. 

\section{Rotation of the jump operators} \label{sec_appx_openr}

In the main text we have introduced $SO(3)$ rotations and corresponding $SU(2)$ transformations that take a generic two-qubit interaction Hamiltonian, $H_{\rm proble}$ to the `diagonal' form $\widetilde{H}_{\rm probe}$. For the noisy case we see that all four coincidence conditions are satisfied only for specific kinds of noise like amplitude damping acting on both qubits. It is important to note that we are introducing amplitude damping noise generated by the $\sigma_-^{(j)}$ operator while simultaneously assuming that the Hamiltonian has been brought to the `diagonal' form. In other words, the jump operator is $\sigma_-^{(j)}$ in the operator basis that make the Hamiltonian equal to $\widetilde{H}_{\rm probe}$. Here we examine what the noise process is in the original operator basis in which the Hamiltonian does not have the 'diagonal' form. As an example, we consider the permutation Hamiltonian,
\[
H_{\rm probe} = g \left(\eta_{xy}\sigma_x \sigma_y + \eta_{yz}\sigma_y \sigma_z + \eta_{zx}\sigma_z \sigma_x \right).
\]
A comparison with the most general form of the probe Hamiltonian given in Eq.~\eqref{inter} allows us to identify the coefficient matrix corresponding to the permutation Hamiltonian as, 
\begin{align}
    \eta = \begin{bmatrix}
        0 && \eta_{xy} && 0\\
        0 && 0 && \eta_{yz} \\
        \eta_{zx} && 0 && 0
    \end{bmatrix}.
\end{align}
A singular value decomposition of the coefficient matrix as \(\eta = U D V^T\) yields,
\begin{align}
    U \! = \! -\!\begin{bmatrix}
        0 & 1 & 0\\
        1 & 0 & 0 \\
        0 & 0 & 1
    \end{bmatrix}, \,
    D \! = \! \begin{bmatrix}
        \eta_{yz} & 0 & 0\\
        0 & \eta_{zy} & 0 \\
        0 & 0 & \eta_{zx}
    \end{bmatrix}, \, 
    V \! = \! - \!\begin{bmatrix}
        0 & 0 & 1\\
        0 & 1 & 0 \\
        1 & 0 & 0
    \end{bmatrix}, \;
\end{align}
with $\det(U)=\det(V)=1$ so they are proper $SO(3)$ rotations and not reflections $(\det = -1)$. Since the SVD is not unique and we can always choose $U$ and $V$ in this manner. We obtain the corresponding local \(SU(2)\) transformations on the $\sigma_j$ operators that equivalently implement the $SO(3)$ transformations $U$ and $V$ as,
\begin{align}
    u_i = \exp\left[-i \frac{\pi}{2\sqrt{2}} \vec{n}_i \cdot \vec{\sigma} \right]
\end{align}
where $\vec{n}_1 = (-1, 1, 0)$ and  $\vec{n}_2 = (-1, 0, 1)$. The transformations we require on the $\vec{\sigma} = \{\sigma_x, \sigma_y, \sigma_z\}$ follow from the singular value decomposition of $\eta$ as $\vec{\sigma}^{(1)} = U\vec{\sigma}$ and $\vec{\sigma}^{(2)} = V\vec{\sigma}$ It is easy to verify that the unitaries, $u_i$ do transform the three Pauli matrices as, 
\begin{eqnarray}
    \sigma_x^{(1)} \rightarrow -\sigma^{(1)}_y, &\;\; \sigma_y^{(1)} \rightarrow -\sigma^{(1)}_x , & \;\; \sigma_z^{(1)} \rightarrow -\sigma^{(1)}_z, \nonumber \\
    \sigma_x^{(2)} \rightarrow -\sigma^{(2)}_z, &\;\; \sigma_y^{(2)} \rightarrow -\sigma^{(2)}_y , & \;\; \sigma_z^{(2)} \rightarrow -\sigma^{(2)}_x,
    \label{eq:sigmatransform}
\end{eqnarray}  
so that 
\[ \widetilde{H}_{\rm probe} = \left[u_1^{\vphantom{\dagger}} \otimes u_2^{\vphantom{\dagger}}\right] H_{\rm probe} \left[u_1^\dagger \otimes u_2^\dagger\right] = g \!\! \sum_{i=x,y,z} \!\! \tilde{\eta}_i\sigma_i \otimes \sigma_i,\]
with $\tilde{\eta}_x = \eta_{yz}$, $\tilde{\eta}_y = \eta_{xy}$ and $\tilde{\eta}_z = \eta_{zx}$.
In other words, we see that the permutation Hamiltonian can be  rotated to become identical to the Hamiltonian of the the anisotropic Heisenberg model and it also reduces to the Hamiltonian of the flip-flop model that we have used extensively if $\eta_{zx}=0$.

One possible initial state that satisfies the four coincidence conditions for the anisotropic Heisenberg model is $|01 \rangle$. In order to find the optimal initial state for the permutation Hamiltonian in its original (non-rotated) operator basis, we can apply the inverse (adjoint) of the local $SU(2)$ transformation as $(u_1^\dagger \otimes u_2^\dagger) |01 \rangle$. Similarly the jump operators $\sigma_- = (\sigma_x - i\sigma_y)/2$ corresponding to amplitude damping on both the qubits in the basis in which the Hamiltonian is `diagonal' correspond to the following operators in the original basis:
\begin{align}
    L_1 &= u_1^\dagger \sigma_-^{(1)} u_1 = \frac{ i \sigma_x^{(1)} - \sigma_y^{(1)} }{2} = iX \sigma_-^{(1)}X^\dagger = i\sigma_+^{(1)}, \nonumber\\
    L_2 &= u_2^\dagger \sigma_-^{(2)} u_2 = \frac{ i \sigma_y^{(2)} - \sigma_z^{(2)}}{2} = -h_{\rm d} \, \sigma_-^{(2)} h_{\rm d},
\end{align}
where $X= e^{-i\frac{\pi}{2}\sigma_x}$ corresponds to a rotation of $\pi$ about the $\sigma_x$ axis, $h_{\rm d}$ is the Hadamard transformation that rotates between the eigenbasis of $\sigma_z$ and $\sigma_z$. This means that in the original operator basis of the permutation Hamiltonian, if all four coincidence criteria are to be simultaneously satisfied, then one possible choice for the dissipators in the master equation is given by \(\sum_{i=1,2} \mathcal{D}\left[L_i\right]\). 
Of course, since $L_1$ and $L_2$ are not the simple lowering operators that would correspond to amplitude loss, via e.g. spontaneous emission, in practice they would not occur naturally.  Nonetheless we can comment on the nature of the dissipation that they correspond to.  
The new dissipators reflect the rotation of the Pauli operator basis given in Eq.~\eqref{eq:sigmatransform}. The amplitude damping channel that we are interested in is a transformation of the Bloch sphere of states of a qubit eventually into a point (sphere of radius zero) located at the south pole or ground state, $|1\rangle$, of the qubit. The transformation in Eq.~\eqref{eq:sigmatransform} takes $\sigma_z$ to $-\sigma_z$ for qubit 1, thereby interchanging the ground and excited states. This is reflected in $L_1$ which is proportional to $\sigma_+$ signifying decay to the excited state $|0\rangle$ corresponding to the flipped $\sigma_z$ axis. For the second qubit the transformation replaces $\sigma_z$ with $-\sigma_x$. This means that in this basis the amplitude damping should be contraction of the Bloch sphere of states to the point on the right end of the $\sigma_x$ axis (the $|+\rangle$ state). The Hadamard transformation, $h_{\rm d}$ exchanges $\sigma_x$ and $\sigma_z$ leaving $\sigma_y$ invariant except for a minus sign and indeed, $L_2$ is related to $\sigma_-^{(2)}$ by the Hadamard transformation.

\bibliography{bibliography}

\end{document}